\documentclass[aps,prd,twocolumn,a4paper,superscriptaddress,10pt,floatfix]{revtex4-2}
\usepackage{graphicx,multirow,latexsym}
\usepackage{hyperref}
\usepackage[british]{babel}
\usepackage{amsmath,amsmath}

\begin{document}
\newcommand{\be}{\begin{equation}}
\newcommand{\ee}{\end{equation}}
\newcommand{\bq}{\begin{eqnarray}}
\newcommand{\eq}{\end{eqnarray}}
\newcommand{\bsq}{\begin{subequations}}
\newcommand{\esq}{\end{subequations}}
\newcommand{\bc}{\begin{center}}
\newcommand{\ec}{\end{center}}
\newcommand{\vos}[1]{{\left<#1\right>}}
\newcommand{\Hub}{\mathcal{H}}
\newcommand{\dd}[1]{\text{d}{#1}}
\renewcommand{\vec}[1]{\bf{#1}}
\newcommand{\matr}[1]{\boldsymbol{#1}}
\newcommand{\dpartial}[2]{\frac{\partial {#1}}{\partial {#2}}}
\newcommand{\ddfrac}[2]{\frac{\dd{#1}}{\dd{#2}}}
\newcommand{\ddpartial}[2]{\frac{\partial^2 {#1}}{\partial {#2}^2}}
\newcommand{\dddfrac}[2]{\frac{\dd{}^2 {#1}}{\dd {#2}^2}}
\newcommand{\dprime}[1]{{#1^\prime}}
\newcommand{\ddprime}[1]{{#1^{\prime\prime}}}

\title{Current-carrying string network evolution in an external magnetic field}
\author{P. B. Barbosa}
\email{up202006852@edu.fc.up.pt}
\affiliation{Centro de Astrof\'{\i}sica da Universidade do Porto, Rua das Estrelas, 4150-762 Porto, Portugal}
\affiliation{Faculdade de Ci\^encias, Universidade do Porto, Rua do Campo Alegre, 4150-007 Porto, Portugal}
\author{C. J. A. P. Martins}
\email{Carlos.Martins@astro.up.pt}
\affiliation{Centro de Astrof\'{\i}sica da Universidade do Porto, Rua das Estrelas, 4150-762 Porto, Portugal}
\affiliation{Instituto de Astrof\'{\i}sica e Ci\^encias do Espa\c co, Universidade do Porto, Rua das Estrelas, 4150-762 Porto, Portugal}
\date{\today}

\begin{abstract}
Cosmic strings are topological defects arising in a variety of cosmological scenarios, as the Universe undergoes symmetry-breaking phase transitions, whose discovery would offer valuable insight into the high-energy physics that shaped the early Universe. To interpret such a detection, robust theoretical models are essential. The Velocity-dependent One-Scale (VOS) model is particularly prominent: it self-consistently treats the network as a thermodynamic system, characterizing its key properties, and predicting its large-scale evolution. This has recently been extended to include superconducting cosmic strings, which carry additional degrees of freedom, giving rise to the Charge-Velocity dependent One-Scale (CVOS) model. One limitation of the latter model is that it only included loss mechanisms for the charge and current. Here, we extend this model by phenomenologically including a possible energy source mechanism, specifically by allowing current-carrying strings interactions with an external magnetic field. We  discuss how this coupling impacts the network's evolution, present and classify the physically allowed scaling solutions for this extended CVOS model, and comment on the different impacts of the external magnetic field on the evolution of the network's charge and current. Under our modeling assumptions, one of the ten physically plausible scaling solution enables the network to gain energy by this mechanism.
\end{abstract}
\maketitle
%%%%%%%%%%%%%%%%%%%%%%%%%%%%%%%%%%%%%%%%%%%%%%%%%%%%%%%%%%%%%%%%%%%%%%%%%
\section{\label{section1}Introduction}

Cosmic strings are compelling relics of symmetry-breaking phase transitions in the early Universe, emerging naturally through the Kibble mechanism \cite{Kibble76,VSbook}. They arise generically within a variety of high-energy physics frameworks, ranging from grand unified theories to string-inspired brane inflation models. Their potential observational signatures render them valuable probes of the early Universe and of fundamental physics. A robust understanding of their cosmological implications, however, relies on a detailed and consistent description of their dynamical evolution which, given their non-linear nature, requires a combination of analytical modeling and numerical analysis. In this work, we employ existing analytical tools to deepen our understanding of the evolution of current-carrying cosmic string networks.

Whereas numerical simulations explore discretized version of exact configurations solving the field equations governing the dynamics of cosmic strings, analytical studies are generally compelled to adopt effective descriptions in which the network is treated as a thermodynamical system. Such an approach was originally proposed by Kibble in \cite{Kibble:1984hp}, where the network was characterized by a single length scale, associated with the total energy density, and allowed to evolve in time. Subsequent developments extended the formalism to include the root mean square velocity as an additional macroscopic parameter \cite{Martins:1995tg,Martins:1996jp,Martins:2000cs,Martins_2016}. This refined framework, known as the velocity-dependent one-scale (VOS) model, has since been calibrated and validated through extensive and statistically robust comparisons with numerical simulations \cite{Correia_2019,Correia:2021tok}.

Although the above framework fully describes the simplest, featureless cosmic string networks, in physically more realistic scenarios the strings are expected to possess internal degrees of freedom, e.g. charges and currents \cite{Witten:1984eb}. In these cases, their observational consequences can change dramatically, the simplest example being the fact that as these strings move through a plasma most of their energy losses will be through electromagnetic radiation (and possibly also particles), with the gravitational radiation contribution being at best subdominant and most likely entirely negligible, depending on the dimensionless string tension $G\mu$ \cite{OTW1,OTW1b,OTW2,OTW3,OTW4}. A subsequent extension enabled the VOS model to account for such generic string networks, incorporating additional degrees of freedom that can be interpreted as effective charges and currents, which moreover can evolve differently---unlike an earlier version, which only addressed the chiral case \cite{Oliveira_2012}. This extension, known as the Charge–Velocity dependent One Scale (CVOS) model, provides a framework for describing the macroscopic evolution of current–carrying cosmic strings \cite{Martins:2020jbq, Martins_2021}. Within this formulation, additional effective characteristic length scales are introduced to capture the impact of both charges and currents on the network’s energy. Efforts to numerically calibrate the CVOS model are ongoing \cite{Correia24}.

One limitation of the aforementioned extension is that it only explicitly includes energy loss mechanisms for the network, from loop production to charge and current losses, although, at a purely phenomenological level, energy gain mechanisms could be accounted for by changing signs on model parameters normalizing these terms (e.g., a negative loop chopping efficiency). Here, we explicitly consider the impact of a possible energy source for the cosmic string network for superconducting strings, specifically  an external magnetic field. Several mechanisms have been proposed for the generation of primordial magnetic fields \cite{Lemoine,Gasperini,Sigl,Trivedi:2013wqa,Subra,Tanmay}, which could permeate the Universe on large scales. On the other hand, we have already observed magnetic fields in galaxies, which are generally attributed to dynamo processes \cite{Miranda:1998ne,Beck,Widrow,Planck}. The effect of a magnetic field on an isolated superconducting loop has already been studied in \cite{Witten:1984eb}. Here, we extend this line of research to the case of string networks by incorporating this mechanism into the CVOS formalism. Once the macroscopic equations incorporating an external magnetic field are derived, we will seek scaling solutions. Similar analyses have previously been carried out for the chiral limit \cite{Oliveira_2012}, for the generalized CVOS equations \cite{Pimenta:2024blk, pimenta2025} as well as for wiggly strings \cite{Almeida:2021ihc,Almeida_2022}.

The outline of the rest of this work is as follows. We start in section \ref{section2}, with a brief introduction to the CVOS model, highlighting its most important features, which will allow us to incorporate the effects of an external magnetic field. Still in \ref{section2}, we clarify the method by which late-time scaling solutions can be obtained from the resulting equations. In \ref{section3}, we analyze these solutions and, where possible, explicitly map their parameter space. Finally, in \ref{section4}, we present our main conclusions. Throughout the work we are using units where the speed of light is $c=1$.

%%%%%%%%%%%%%%%%%%%%%%%%%%%%%%%%%%%%%%%%%%%%%%%%%%%%%%%%%%%%%%%%%%%%%%%%%
\section{\label{section2}The CVOS Model and its extension}

Here we briefly introduce the CVOS formalism, following the original work in \cite{Martins:2020jbq, Martins_2021}. The model describes the macroscopic evolution of the string network, starting with the action:
\begin{equation}
    S = -\mu_0 \int \sqrt{-\gamma} f\left( k\right) \,\,d^2\sigma\,,
\end{equation}
where $\mu_0$ is the string tension, $\gamma$ is the determinant of the string's worldsheet metric, given by $\gamma_{ab} = g_{\mu \nu} x^\mu_{,a} x^\nu_{,b}$, $x^{\mu}$ denoting the coordinates of the position 4-vector and $k$ is the chirality,
\begin{equation}
    k = \gamma^{ab} \phi_{,a} \phi_{,b} = \frac{\dot{\phi}^2}{a^2 \left(1-\dot{\mathrm{x}}^2\right)} - \frac{\phi'^2}{a^2 \mathrm{x}'^2} = q^2 - j^2,
\end{equation}
where $\phi$ is the scalar field, defined solely on the strings, responsible for generating charges and currents. These are defined above and correspond to $q$ and $j$, respectively. The model allows for an expanding Universe, by considering a background metric, which we assume to be the Friedmann-Lema\^{\i}tre-Robertson-Walker one. It can be observed that the microscopic charge and current depend on the expansion of the Universe through the scale factor $a\left(\tau\right)$. The dot and prime derivatives of the string 3-coordinates are with respect with the string worldsheet coordinates, $\tau$ and $\sigma$, respectively. From this action the microscopic equations can be deduced, and conveniently written in terms of the following three quantities
\begin{subequations}
    \begin{align}
        \tilde{U} &= f - 2q^2\frac{df}{dk}, \\
        \tilde{T} &= f + 2j^2\frac{df}{dk}, \\
        \Phi &= -2qj\frac{df}{dk}\,.
    \end{align}
\end{subequations}

In order to derive the macroscopic quantities from the microscopic ones, the model relies on averaging through the entire network. The average of any quantity $\mathcal{O}$ is given by
\begin{equation}\label{averageq}
    \langle \mathcal{O} \rangle = \frac{\int \mathcal{O} \epsilon \, d\sigma}{\int \epsilon \, d\sigma},
\end{equation}
where $\epsilon = \sqrt{\frac{\mathrm{x}'^2}{1-\dot{\mathrm{x}}^2}}$ is the dimensionless bare energy density. In this framework, we can determine the energy related solely to the strings, called the bare string energy, $E_0$, and the total energy, which also includes the charge and current contribution, $E$. They are defined as $E_0 = \int \epsilon \, d\sigma$ and $E = \int \tilde{U} \epsilon \, d\sigma$, respectively. From these two definitions, we can determine the two characteristic length scales, each related to the corresponding energy: $E_0 = \frac{\mu_0 V}{a^2 \xi^2}$ and $E = \frac{\mu_0 V}{a^2 L^2}$, where $\xi$ and $L$ are henceforth referred to as the correlation and characteristic lengths, respectively. We can further define the macroscopic root-mean square velocity, $v$, charge, $Q$, and current, $J$, through the average of their microscopic counterparts.

By  this averaging process, we can obtain the evolution of the relevant macroscopic quantities. However, this process does not account for energy losses such as the production of loops. The introduction of further phenomenological terms to characterize these energy losses is well documented in \cite{Martins_2016} and was later extended to CVOS in \cite{Martins:2020jbq}. Taking these into consideration, we get
\begin{subequations}
	\begin{align}
		\dot{L} &= \frac{\dot{a}}{a} \frac{L}{W^2}\left\{ \left[W^2+\left(Q^2+J^2\right)F'\right]v^2 - \left(Q^2+J^2\right) F'\right\}\\        &+ g\frac{\tilde{c}}{2}\frac{v}{W}\notag \\
		\dot{v} &= \frac{1-v^2}{W^2} \left\{\frac{k_v}{\xi}\left[W^2 - 2\left(Q^2+J^2\right)F'\right]\right\} \\ &- \frac{1-v^2}{W^2}\left\{2v\frac{\dot{a}}{a}\left[W^2+\left(Q^2+J^2\right)F'\right]\right\} \notag\\
		\frac{d J^2}{d \tau} &= 2 J^2 \left[\frac{v k_v}{\xi} - \frac{\dot{a}}{a}\right] + \rho \tilde{c}\frac{v}{L}\frac{\left(g-1\right)W}{F'-2Q^2F''}\\
		\frac{d Q^2}{d \tau} &= 2 Q^2 \frac{F' + 2J^2 F''}{F' + 2Q^2 F''}\left[\frac{v k_v}{\xi} - \frac{\dot{a}}{a}\right]\\ &+ \left(1-\rho\right) \tilde{c}\frac{v}{L}\frac{\left(g-1\right)W}{F'+2Q^2F''}\notag.
	\end{align}
\end{subequations}

Here $F = \langle f \rangle$, $F' = \langle \frac{df}{dk} \rangle$ and $F'' = \langle \frac{d^2f}{dk^2} \rangle$. $k_v$ is the momentum parameter and $W = \sqrt{F - 2Q^2F'}$ relates both length scales, specifically through $\xi = W L$. As for the phenomenological energy loss parameters, $\tilde{c}$ is the loop chopping efficiency. The parameter $g$ is the overall loop production bias parameter, measuring whether regions with charge and current are more or less prone to be involved in loop production. If $g>1$, then regions with these degrees of freedom will have a higher contribution to loop production. These loops will retain these charges and currents leading to a general loss on its energy contribution for the overall string network. This would result in smaller values of $Q$ and $J$. So, instead of interpreting the $g$ parameter in terms of the role of spatial variations of the microscopic charge and current on the likelihood of loop production, we can interpret it through a more averaged point of view. Specifically, we can link the value of $g$ to the readiness on the averaged charge and current to vary in time. If $g>1$, then loop production will lead to smaller values of $Q$ and $J$; if $g<1$, then the opposite happens. The case when loop production does not influence the value of charge and current corresponds to $g=1$.

Finally, $\rho$ is the specific charge-current bias parameter. Extending the time variation interpretation, we can view $\rho$ as measuring how distinctively the charge and the current vary with the production of loops. In the unbiased case, $\rho=\frac{1}{2}$, both $Q$ and $J$ will be affected in the same way; if $\rho \rightarrow 0$, then the charge will be highly sensitive to loop production, while the current will feel no repercussions; if $\rho \rightarrow 1$, then the opposite happens. Although the spatial variation interpretation of the parameter $g$ can be extended to the bias parameter, it raises some difficulties in explaining some features in the $g<1$ regime. Such impediments do not occur for the time variation interpretation and, therefore, we shall continue with it. It is also coherent with the features of having $g=1$. In this regime $\rho$ is ill-defined, as it should be expected in a situation where the charge and the current are not influenced a priori by loop production.

\subsection{Including a Magnetic Field}

We can allow for an explicit external magnetic field through the implementation of a coupling term between the scalar field and the gauge fields, relying on the action introduced in \cite{Witten:1984eb}
\begin{equation}
\label{Witten_action}
    S = -\mu_0 \int \left\{ \sqrt{-\gamma} - \frac{1}{2}\sqrt{-\gamma}\gamma^{ab} \phi_{,a} \phi_{,b} + e A_\mu x^\mu_{,a} \varepsilon^{ab} \phi_{,b} \right\} d^2 \sigma,
\end{equation}
where $e$ is the coupling constant and $A_\mu$ is the gauge field.

Note that for simplicity we are taking the linear case $f=1-\frac{1}{2}k$, as in \cite{Martins_2021}. This is expected to be an adequate description in most astrophysical contexts, for example is is known that in the Witten superconducting model, and for almost chiral currents, the network evolution reduces dynamically to that of the linear case \cite{WittenModel}. Nevertheless, a possible exception would be scenarios with very large (or saturated) charges or currents, in which case non-linear effects may become relevant. However, in the present context we will see that this is only the case in one particular solution, i.e. in a small fraction of the model's parameter space. With that caveat, the relation between both length scales becomes well determined, with $W=\sqrt{1+\frac{1}{2}\left(Q^2+J^2\right)}$. 

Following \cite{Tsagas_2004}, it is possible to define the magnetic field $\mathbf{B}$ through $F_{ij} = -a^2 \tilde{\varepsilon}_{0ijk} \delta^{kl} B_l$, where $B_l$ are just the coordinates of the 3-vector $\mathbf{B}$ and $\tilde{\varepsilon}_{\mu \nu \rho \lambda}$ is the Levi-Civita pseudo-tensor, with $\tilde{\varepsilon}_{0123} = 1$. It should be noted that the coupling term also allows the charges and currents on the strings to generate their own electromagnetic fields. However, given their small magnitudes, their influence on the string dynamics is overwhelmingly dominated by the external magnetic field. Therefore, in the present work, we neglect these self-generated fields. Within the CVOS formalism, the averaged equations of motion resulting from the previous action become
\begin{subequations}
\label{Equations_CVOS_Magnetic}
	\begin{align}
		\dot{L} &= \frac{\dot{a}}{a} \frac{L}{W^2}\left\{ \left[W^2-\frac{1}{2}\left(Q^2+J^2\right)\right]v^2 + \frac{1}{2} \left(Q^2+J^2\right)\right\}\\  &+ g\frac{\tilde{c}}{2}\frac{v}{W} \notag\\
		\dot{v} &= \frac{1-v^2}{W^2} \left\{\frac{k_v}{\xi}\left[W^2 +\left(Q^2+J^2\right)\right]\right\}\\ &-\frac{1-v^2}{W^2}\left\{2v\frac{\dot{a}}{a}\left[W^2-\frac{1}{2}\left(Q^2+J^2\right)\right] - e a B Q \theta\right\} \notag\\
		\frac{d J^2}{d \tau} &= 2 J^2 \left[\frac{v k_v}{\xi} - \frac{\dot{a}}{a}\right] -2 \rho \tilde{c}\frac{v}{L}\left(g-1\right)W\\
		\frac{d Q^2}{d \tau} &= 2 Q^2 \left[\frac{v k_v}{\xi} - \frac{\dot{a}}{a}\right]-2\left(1-\rho\right) \tilde{c}\frac{v}{L}\left(g-1\right)W\\ &-2e a B v Q \theta\,.\notag
	\end{align}
\end{subequations}
Two new quantities related to the magnetic field have been introduced: the effective mean intensity of the magnetic field, $B$, defined through $\langle \mathbf{B} \rangle = B \langle \hat{n} \rangle$, where $\hat{n}$ is a unit vector pointing in the direction of $\mathbf{B}$, and the parameter $\theta = \langle\frac{\left( \dot{\mathbf{x}} \times \hat{n} \right) \cdot \mathbf{x}'}{\sqrt{\dot{\mathrm{x}}^2 \mathrm{x'^2}}} \rangle$, which quantifies the relative orientation between the string direction, its velocity and the magnetic field.

This $\theta$ parameter is a dimensionless quantity ranging from $-1$ to $1$ and resembles the Lorentz force in the string direction. Locally, it implies that an external magnetic field influences the string current only if it possesses a non-vanishing component perpendicular to the plane spanned by $\mathbf{\dot{x}}$ and $\mathbf{x'}$. Indeed, in the simple case of an asymptotically Nambu-Goto circular loop (i.e., one with a tiny charge and current) in a magnetic field that is perpendicular to the loop's plane one would have $\theta=\pm1$ depending on the magnetic field's orientation, while if the magnetic field orientation is in the loop's plane, $\theta=0$. This means that at the microscopic level and for a perfectly isotropic string network theta would vanish. However, in the VOS model context one considers a mesoscopic average, not a microscopic one. An instructive analogy can be drawn with the renormalized string mass per unit length, $\tilde\mu$, in wiggly string models \cite{Wiggly}. Locally, $\tilde\mu$ is identically equal to unity, but when defined as a mesoscopic quantity, aimed at characterizing small-scale structure at a coarse-graining scale suitable for macroscopic models such as VOS, it generally satisfies $\tilde\mu>1$. A similar mechanism should hold in the present case: $\theta$ may vary on length scales smaller than the correlation length, driven by variations in the string velocity and direction along the worldsheet. As a result, neighboring regions in which $\theta$ changes are not entirely uncorrelated, opening the possibility for a non-vanishing averaged value of $\theta$ at the network level. For a generic current-carrying string network the averaged value of $\theta$ could be positive or negative, but there is in principle no expectation that this average, which is defined by Eq. (\ref{averageq}), should vanish, especially in the biased cases. Ultimately, its value will need to be determined from numerical simulations.

For sake of simplicity $\theta$ is assumed to be a constant in the next sections of the paper. However, in order to gain a better understanding of its possible value, we note that it can also be treated as an evolving parameter and one can determine its evolution equation. In fact, we have
\begin{align}
\label{theta_evolution}
    \frac{d\theta}{d\tau} &= \frac{\sqrt{1-v^2}}{W^2}JB\theta_0 \left[ 2\frac{Q}{v}k_1 + eaB\theta_1\right] + Bk_1\theta_1 \\ 
    &+ \frac{1-v^2}{v}\frac{B}{W^2}\{ \left[ W^2 - \left(Q^2+J^2\right)\right] k_2 \theta_0 + eaQB\nu \}, \notag
\end{align}
where several new parameters have been defined
\begin{subequations}
    \begin{align}
    k_1 &= \biggr \langle \frac{\dot{\mathbf{x}}' \cdot \hat{e}_2}{\sqrt{\mathrm{x}'^2}}\biggr\rangle \\
    k_2 &= \biggr \langle \frac{\mathbf{x}'' \cdot \hat{e}_2}{\mathrm{x}'^2}\biggr\rangle \\
    \theta_0 &= \frac{1}{B} \langle \mathbf{B} \cdot \hat{e}_0\rangle \\
    \theta_1 &= \frac{1}{B} \langle \mathbf{B} \cdot \hat{e}_1\rangle \\
    \nu &= \frac{1}{B^2} \langle \left(\mathbf{B} \cdot \hat{e}_0\right)^2\rangle. 
\end{align}
\end{subequations}
We have also introduced the trihedron $(\hat{e}_0, \hat{e}_1, \hat{e}_2)$, which is defined in each point of the string network and consists of three orthogonal unit vectors pointing along $\dot{\mathbf{x}}$, $\mathbf{x}'$ and $\dot{\mathbf{x}} \times \mathbf{x}'$, respectively. A more detailed explanation of this trihedron and of the deduction of  Eq.\ref{theta_evolution}, as well as the evolution equations for the parameters $\theta_0$, $\theta_1$ and $\nu$ can be seen in Appendix \ref{app1}.

The first point to notice regarding Eq. \ref{theta_evolution} is that its right-hand side vanishes if there is no magnetic field. It is not a trivial equation and it is not straightforward to infer how $\theta$ might evolve. In particular, it is clear that the coherence scale of the magnetic field will be relevant. Nevertheless, even if one assumes, from symmetric reasons, that the $\theta_i$ should vanish, there is still a non-vanishing term involving $\nu$, which is a positive definite quantity. In summary, while we currently cannot quantitatively estimate the value of $\theta$ (which would require numerical simulations), there is no reason to expect that it vanishes.

We note that only the evolution equations for the velocity and charge have gained explicit magnetic field dependent terms. Moreover, comparing the evolution equations for charge and current, we note that in the absence of the primordial magnetic field for the unbiased case $\rho=1/2$ the two evolution equations would be identical, corresponding to the chiral case \cite{Oliveira_2012,Martins:2020jbq}; clearly, the presence of the magnetic field breaks this symmetry.

\subsection{Scaling solutions}

The CVOS dynamical equations can evidently be solved numerically for any cosmological model of choice. Still, much insight can be gained by searching for and classifying analytical solutions in specific cosmological epochs, including asymptotic solutions, which can be tested against numerical simulations \cite{Correia24}. For featureless cosmic string networks the attractor solution is the well-known linear scaling, in which the correlation length is proportional to the horizon and the velocity remains constant. In models with additional degrees of freedom, additional solutions are possible, depending on the cosmological epoch and model parameter values. A detailed classification of these solutions for wiggly strings can be found in \cite{Almeida:2021ihc,Almeida_2022}, while for the CVOS model it is reported in \cite{Pimenta:2024blk, pimenta2025}. Here, we will revisit the latter work, with the addition of a primordial magnetic field. We provide a classification of these scaling solutions in the following section, but here we briefly outline the physico-mathematical procedure by which they can be obtained.

We are looking for asymptotic solutions in which the macroscopic quantities scale as a power of the conformal time
\begin{subequations}
	\begin{align}
		L &= L_0 \tau^{\alpha} \\
		v &= v_0 \tau^{\beta} \\
		Q &= Q_0 \tau^{\gamma} \\
		J &= J_0 \tau^{\delta} \\
		\xi &= \xi_0 \tau^{\epsilon}.
	\end{align}
\end{subequations}
Importantly, not all mathematically allowed solutions are physically acceptable, implying that there are some constraints on the exponents, which exclude a significant fraction of otherwise allowed cases.

The characteristic length cannot be larger than the correlation length, which cannot be larger than the Hubble radius. This implies that $\alpha \leq \epsilon \leq 1$. On the other hand, the velocity cannot be greater than $1$, implying that $\beta$ cannot be positive: $\beta \leq 0$. The above ansatz is particularly useful when the scale factor also scales as a power law in conformal time, as it is the case for the radiation era, $a \propto \tau$, and the matter era, $a \propto \tau^2$. In what follows, we will assume the generic power law form $a = a_0 \tau^\lambda$, occasionally considering the specific radiation or matter cases.

Last but not least, in our case the magnetic field is also allowed to vary in time. Introducing the conformal magnetic field as in \cite{Tanmay}, we will also write it in power law form
\be
B_c = a^2 B \propto\tau^p\,,
\ee
where $p$ is an additional exponent which, as will be seen, will impact scaling solutions. On the other hand, in the present work we will ignore its spatial dependence, effectively treating it as an averaged (coarse-grained) magnetic field, on a suitably large scale which in the context of the CVOS model does not be explicitly specified. We do note that this scale need not be that of the string correlation length itself---all that is required is that this averaging can be done on some mesoscopic scale. In practice we are treating this time-dependent magnetic field in an analogous way to the case of the purely time-dependent renormalized string mass per unit length in the case of wiggly strings, a discussion on which can be found in \cite{Wiggly}.

Since the charges and currents carried by the strings are assumed to be sufficiently small to justify a linear dependence of the function $f$ on the chirality, the electromagnetic field generated by the string is expected to be subdominant relative to the background magnetic field, and electromagnetic backreaction, although not exactly zero, is expected to be correspondingly small, and has therefore been neglected.

By substituting the above power laws in Eqs. \ref{Equations_CVOS_Magnetic}, we get the following set of equations
\begin{subequations}
	\label{scaling_equations_full}
	\begin{align}
		\alpha &= \lambda \left[v_0^2 \tau^{2\beta} + \frac{1}{2}\frac{L_0^2}{\xi_0^2}\mathcal{D}_0^2 C_v \tau^{2\left(\alpha-\epsilon+\eta\right)}\right] + \frac{g \tilde{c}}{2} \frac{v_0}{\xi_0} \tau^{1+\beta-\epsilon} \\
		\frac{\beta v_0}{C_v} &=  \frac{k_v}{\xi_0}\left[1-\frac{L_0^2}{\xi_0^2}\mathcal{D}_0\tau^{2\left(\alpha-\epsilon+\eta\right)}\right] \tau^{1-\beta-\epsilon}\\ &- 2v_0 \lambda \left[1-\frac{1}{2}\frac{L_0^2}{\xi_0^2}\mathcal{D}_0\tau^{2\left(\alpha-\epsilon+\eta\right)}\right]\notag\\ &+ \frac{L_0^2}{\xi_0^2}Q_0 \Delta \tau^{2\alpha-2\epsilon+1+\gamma-\lambda-\beta+p}\notag \\
		\delta &= \frac{v_0 k_v}{\xi_0}\tau^{1+\beta-\epsilon}-\lambda-\rho\left(g-1\right)\tilde{c}\frac{v_0 \xi_0}{J_0^2 L_0^2}\tau^{1+\beta+\epsilon-2\alpha-2\delta} \\
		\gamma &= \frac{v_0 k_v}{\xi_0}\tau^{1+\beta-\epsilon}-\lambda\\ &-\left(1-\rho\right)\left(g-1\right)\tilde{c}\frac{v_0 \xi_0}{Q_0^2 L_0^2}\tau^{1+\beta+\epsilon-2\alpha-2\gamma}\\ &-\frac{v_0}{Q_0}\Delta \tau^{1+\beta-\gamma-\lambda+p}.\notag
	\end{align} 
\end{subequations}
Several new parameters have been introduced for convenience. In order to understand them we should bear in mind that our ansatz aims to characterize the network at late times, meaning that we effectively consider $\tau \rightarrow +\infty$. For example, $C_v = 1-v^2 = 1 - v_0^2 \tau^{2\beta}$. If $\beta = 0$, then $C_v$ is a constant. Alternatively, if $\beta <0$, then in the limit $\tau \rightarrow \infty$, $C_v \rightarrow 1$, which is also a constant. Analogously, we define $Q^2+J^2 = \mathcal{D}_0^2 \tau^{2\eta}$, where
\begin{equation}
	\mathcal{D}_0^2 = \left\{ \begin{matrix} Q_0^2, & \mbox{if }\gamma > \delta\\ J_0^2, & \mbox{if }\gamma < \delta\\
		Q_0^2+J_0^2, & \mbox{if }\gamma = \delta. \end{matrix} \right. 
\end{equation}
and the exponent $\eta$ is the largest one between $\gamma$ and $\delta$. 

The parameter $\Delta$ is related to the magnetic field and is particularly useful, since the four unknown parameters $a_0$, $e$, $B_{c,0}$ and $\theta$ always appear in the same way, allowing us to gather them in a single undetermined parameter
\begin{equation}
    \Delta = \frac{e B_{c,0} \theta}{a_0}\,;
\end{equation}
the case where $\Delta=0$ is of secondary interest since this case has already been studied in the literature.

There is one more constraint to be satisfied. It stems from the relation between both length scales and it is therefore related to the ratio between the bare string energy and the total one. This ratio reads
\begin{equation}
	\frac{E_0}{E} = \frac{L_0^2}{\xi_0^2}\tau^{2\left(\alpha-\epsilon\right)} = \frac{2}{2+\mathcal{D}_0^2 \tau^{2\eta}}\,,
	\label{final_restriction}
\end{equation}
and physically three possibilities are allowed, depending on the value of $\eta$
\begin{itemize}
	\item If $\eta = 0$, then we must have $\alpha=\epsilon$, which implies $\frac{L_0^2}{\xi_0^2} = \frac{2}{2+\mathcal{D}_0^2}$;
	\item If $\eta<0$, then we must have $\alpha=\epsilon$ and $\frac{L_0^2}{\xi_0^2} = 1$;
	\item If $\eta>0$, then we must have $\alpha-\epsilon+\eta=0$ and $\frac{L_0^2}{\xi_0^2} \mathcal{D}_0^2 = 2$.
\end{itemize}
As will be seen in what follows, these correspond to the cases of constant, decaying or growing charge and/or current.

\section{\label{section3}Scaling Solutions}

Here we will describe all the possible and physically meaningful solutions stemming from the aforementioned analysis. (Other mathematically allowed but unphysical solutions, such as those with $v=1$, will not be described.) As has already been said, the solutions will fall into three categories, depending on the value of $\eta$: the first family consists of solutions where the dominant degree of freedom---charge or current, depending on whether $\eta=\gamma$ or $\eta=\delta$, respectively---does not decay and $\beta=0$, yielding a constant velocity; the second family consists in solutions where both charge and current decay over time and $\beta = 0$; finally, the third family yields decaying velocity solutions with $\beta<0$ and $\eta=0$.

When numerically exploring the parameter space of specific solutions, we will consider, for simplicity (i.e. with the goal of reducing the number of free parameters), specific values for $\tilde{c}$ and $k_v$. From \cite{Martins:2000cs}, we take $\tilde{c} = 0.23$ and we will consider a constant momentum parameter $k_v$, with the averaged value obtained in \cite{Correia_2019} for a null velocity, $k_0 = 1.32$. 

\subsection{Solutions with $\eta \geq 0$ and $\beta=0$}

In this first family of solutions, there are three different branches, depending on whether the dominant degree of freedom is the charge, the current or both contribute equally. For later use we denote this as case $A$ and the first of its solution branches as $A1$. In this solution, the dominant degree of freedom is the charge which, for the previous values of $\tilde{c}$ and $k_v$, remains constant, while the current decreases. Specifically we have
	\begin{subequations}
		\begin{align}
			\alpha &= \epsilon = 1 \\
			\gamma &= 0 \\
			\delta &= \frac{v_0}{Q_0}\Delta+ \left(g-1\right)\frac{v_0 \xi_0}{Q_0^2 L_0^2} < 0, 
		\end{align}
	\end{subequations}
with the following constrains on the model parameters
\begin{subequations}
		\begin{align}
        &\rho = 0 \\
			&\frac{g\tilde{c}}{2}\frac{v_0}{\xi_0} + \lambda\left[v_0^2+\frac{Q_0^2}{2+Q_0^2}\left(1-v_0^2\right)\right] = 1 \\
			&v_0 \frac{k_v}{\xi_0} - \left(g-1\right)\tilde{c}\frac{Q_0^2}{2+Q_0^2}\frac{v_0}{\xi_0} - v_0\frac{\Delta}{Q_0} = \lambda \\
			&\frac{k_v}{\xi_0}\left(2-Q_0^2\right) + 2Q_0 \Delta = 4v_0 \lambda\,.
		\end{align}
	\end{subequations}
Note that in order to have $\delta<0$ one must either have $g<1$ or $\Delta>0$.
    
One can study the system for different values of $\lambda$, however, for representative purposes, we restrict ourselves to the radiation and matter dominated eras, with $\lambda = 1$ and $\lambda=2$, respectively. It turns out that $A1$ is only possible in the matter era. Considering $-2<\Delta<2$ and $0<g<2$, we solved for the three parameters, $v_0$, $Q_0$ and $\xi_0$, and found that $\frac{v_0}{\xi_0} \approx 1.3$. Using the second restriction one can approximately  write the current exponent as $\delta = v_0 \frac{k_v}{\xi_0} - \lambda \approx -0.3$, meaning a very well restrained current evolution.

\begin{figure}
		\centering
		\includegraphics[width=\columnwidth]{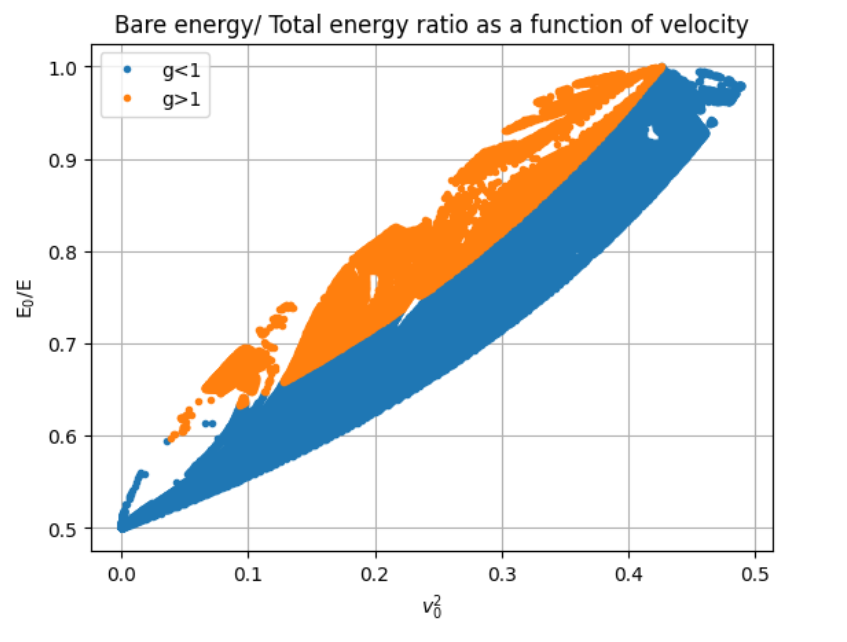}
		\caption{Ratio between the bare string energy $E_0$, and the total energy, $E$, as a function of the final velocity squared, $v_0^2$, for solution $A1$ in the matter era. Points in orange were determined for $g>1$, while blue points correspond to $g<1$.}
		\label{fig:energy_vs_velocity_A1}
	\end{figure}

By knowing the value of $Q_0$ it is also possible to determine the fraction of the total energy on the bare strings themselves (when not considering charges and currents). That value is represented as a function of the velocity in \autoref{fig:energy_vs_velocity_A1}. As we can see, and as is to be expected from \cite{Martins:2020jbq}, networks approaching the Nambu-Goto limit have higher velocities. We also note qualitatively different features depending on whether the parameter $g$ is greater or smaller than $1$, although the overall behavior remains the same. For a given velocity, the energy fraction considering $g>1$ is always greater. This is to be expected because, in such regime, the averaged charge value will decrease, due to loop production, which means, in turn, a smaller energy contribution for this degree of freedom.

In the second solution, henceforth denoted $A2$, the current dominates over the charge, and therefore $\delta>\gamma$. A growing current is not asymptotically allowed: it is physically expected to saturate at some point. Additionally, in the context of the CVOS model, it would mean that the momentum parameter should be $k_v=0$, in direct contradiction with the values we take a priori for the analysis. Therefore, the current must remain constant, and we have
\begin{subequations}
		\begin{align}
			\alpha &= \epsilon = 1 \\
			\delta &= 0 \\
			\gamma &= -\frac{v_0}{Q_0}\Delta + \left(g-1\right)\frac{v_0 \xi_0}{J_0^2 L_0^2} < 0, 
		\end{align}
	\end{subequations}
subject to the restrictions
\begin{subequations}
		\begin{align}
                &\rho = 1 \\
			&\frac{g\tilde{c}}{2}\frac{v_0}{\xi_0} = 1-\lambda+\frac{2}{2+J_0^2}\left(1-v_0^2\right)\lambda\\
			&\frac{k_v}{\xi_0}\left(2-J_0^2\right) = 4v_0 \lambda\\
			&v_0\frac{k_v}{\xi_0} = \lambda+\left(g-1\right)\frac{\tilde{c}}{2}\frac{v_0}{\xi_0}\frac{2+J_0^2}{J_0^2}\\
			&p<\lambda-1.
		\end{align}
	\end{subequations}
In this case, $\gamma<0$ requires $g<1$ or $\Delta>0$.

\begin{figure}
		\centering
		\includegraphics[width=\columnwidth]{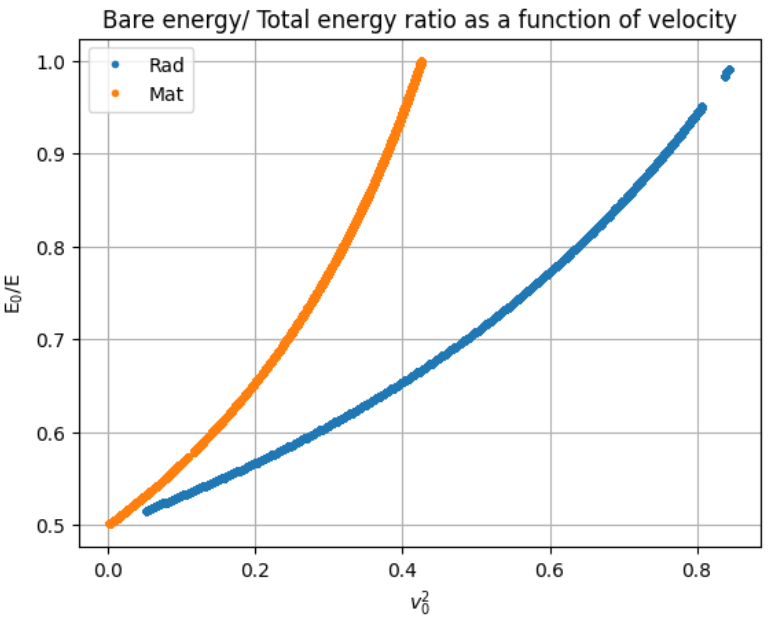}
		\caption{Ratio between the bare string energy $E_0$, and the total energy, $E$, as a function of the final velocity squared, $v_0^2$, in the radiation (blue dots) and matter (orange dots) eras, considering solution $A2$.}
		\label{fig:Energy_vs_velocity_solA2}
	\end{figure}

Again, the values of $v_0$, $J_0$ and $\xi_0$ were obtained for the radiation and matter dominated eras by varying $g$ and $\frac{\Delta}{Q_0}$. The results for both eras are similar, with $\frac{v_0}{\xi_0}\approx 1.3$ and $-2.9 < \gamma < 0$. In this solution the evolution of the necessarily decaying charge has a broader range. By plotting the energy ratio as a function of velocity, we see in \autoref{fig:Energy_vs_velocity_solA2} a behavior analogous to that of solution $A1$, although the values are now far more constrained for a given velocity. There is also a clear difference in the velocity for the two different cosmological epochs. The lower velocities in the matter era are a reflection of a higher Hubble damping.

It should be noted that, in opposition to what happens in solution $A1$, for a given expansion rate only one of the two regimes, $g<1$ or $g>1$, is allowed. More specifically, $g>1$ in the radiation era and $g<1$ in the matter one. There is a critical expansion rate, $\lambda_c$, for which the transition occurs, and it is given by $\lambda_c = \frac{2k_v}{k_v+\tilde{c}}\approx 1.7$. This feature has an intuitive explanation. For higher expansion rates, it is expected for the current decay more easily. So, in order for the current to remain constant, there must be an additional positive feedback. This is what happens when considering the $g<1$ regime, which is precisely in agreement with Eq. \eqref{Equations_CVOS_Magnetic}. The same happens in solution $A1$, for the charge. However, the extra positive feedback results from a joint contribution of loop production and the external magnetic field leading to the mixture of both the $g$ regimes for a given $\lambda$ and possibly causing the broadness on the energy fraction values. 

Comparing the two solutions, we note that the behavior of the characteristic length, correlation length and velocity are akin to those of the linear scaling solution ($\alpha=\epsilon=1$, $\beta=0$), with the magnetic field supplying the energy for the additional degrees of freedom, and the bias parameter $\rho$ determining whether this energy is predominantly retained in the charge or the current.

The third and last solution branch of this family, $A3$, corresponds to the case of identical time dependencies for the charge and the current, $\gamma = \delta$. This solution can be further subdivided into two, depending on whether the charge and current remain constant or grow with time. If $\gamma=\delta=0$, then we must have
\be
p=\lambda-1\,,
\ee
or equivalently $B_c\propto a/\tau$, together with $\alpha=\epsilon=1$, and subject to the following restrictions
\begin{subequations}
\label{restrictions_A3}
		\begin{align}
			&\frac{2}{2+Q_0^2+J_0^2}\left(1-v_0^2\right)\lambda-\frac{g\tilde{c}}{2}\frac{v_0}{\xi_0} = \lambda-1\\
			&\frac{k_v}{\xi_0}\left[2-\left(Q_0^2+J_0^2\right)\right]-4v_0\lambda+2Q_0\Delta=0\\
			&v_0\frac{k_v}{\xi_0}-\rho\left(g-1\right)\tilde{c}\frac{v_0}{\xi_0}\frac{2+Q_0^2+J_0^2}{2 J_0^2} = \lambda\\
			&v_0\frac{\Delta}{Q_0} = \left[\frac{\rho}{J_0^2} - \frac{1-\rho}{Q_0^2}\right] \left(g-1\right)\tilde{c}\frac{v_0}{\xi_0}\frac{2+Q_0^2+J_0^2}{2}.
		\end{align}
	\end{subequations}
    
In this case, the bias parameter $\rho$ is undetermined, making it possible to vary it and determine its influence on the solutions. It should be noted that, similar to solutions $A1$ and $A2$, for all values of $\rho$ we have $\frac{v_0}{\xi_0}=\frac{2}{k_v+\tilde{c}}\approx 1.3$. This raises more restrictions that are not seen at first sight. Specifically, for $\rho=0$, only the critical expansion rate, $\lambda_c=\frac{2 k_v}{k_v+\tilde{c}}$ is possible. The same does not happen in the opposite regime, when $\rho=1$. Considering CVOS without energy loss terms due to loop formation, then this critical expansion rate matches the only way of securing a constant charge and current. Admitting the formation of loops with $\rho=0$, it would mean no difference for the current, because it would feel no repercussions. Therefore, the only way to have a constant $J$, is for the expansion rate to correspond to $\lambda_c$. This does not happen when $\rho=1$, for obvious reasons. In this extended CVOS model, the external magnetic field can cancel any positive or negative increment on the averaged charge, resulting from the expansion of the Universe.

\begin{figure}
		\centering
		\includegraphics[width=\columnwidth]{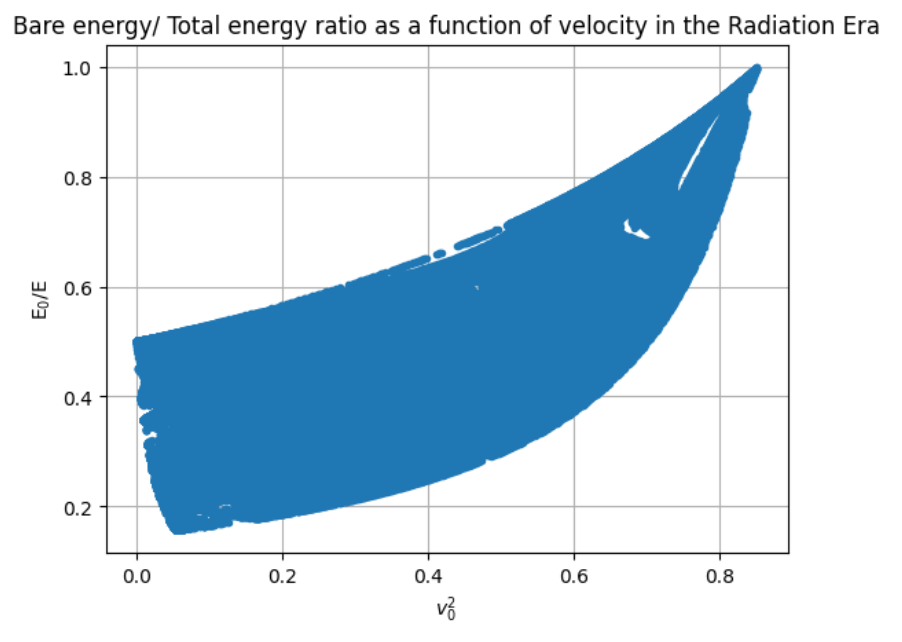}
		\caption{Ratio between the bare string energy $E_0$, and the total energy, $E$, as a function of the final velocity squared, $v_0^2$, in the radiation era, for $\rho=1$, considering solution $A3$ when $\eta=0$.}
		\label{figure3}
	\end{figure}

\begin{figure}
		\centering
		\includegraphics[width=\columnwidth]{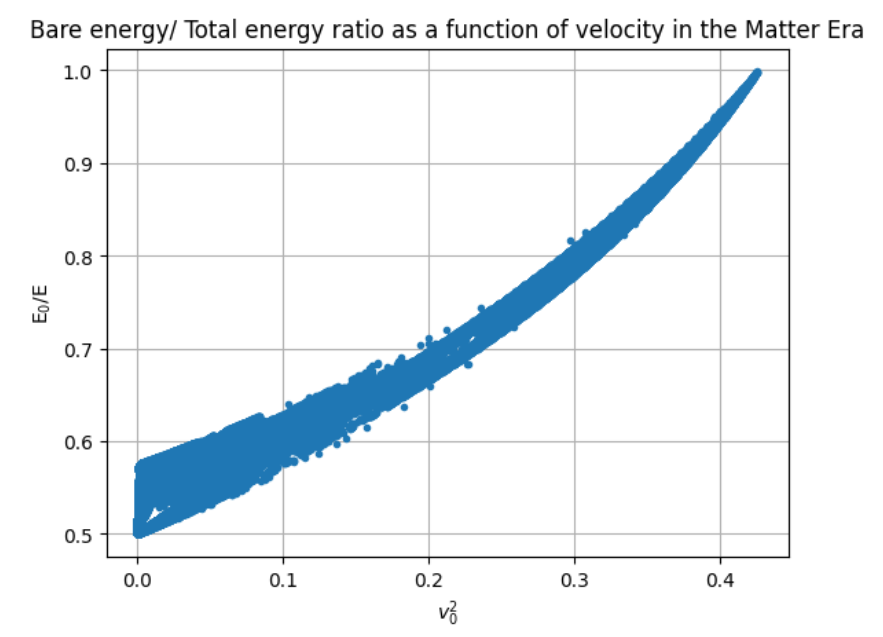}
		\caption{Ratio between the bare string energy $E_0$, and the total energy, $E$, as a function of the final velocity squared, $v_0^2$, in the matter era, for $\rho=1$, considering solution $A3$ when $\eta=0$.}
		\label{figure4}
	\end{figure}

Continuing with the analysis procedure of this family, we shall study how this solution behaves in the radiation and matter dominated eras. Since the value for the critical expansion rate remains the same, we still have $\lambda>\lambda_c \implies g<1$ and vice-versa. The energy fraction values considering $\rho=1$ are represented in \autoref{figure3} and \autoref{figure4}, for $\lambda=1$ and $\lambda=2$, respectively. Again, it is possible to verify that smaller velocities are obtained for higher expansion rates. We can also assert that, in the limit $v_0^2 \rightarrow 0$, the energy fraction approaches $1/2$. The main difference between both expansion rates lies on the width of the possible energy fraction values. We can also imagine a line connecting the points where $\frac{E_0}{E}=\frac{1}{2}$ and $\frac{E_0}{E}=1$ in both figures: it seems that for smaller expansion rates, the energy fraction values tend to lie beneath that line, with the opposite happening for higher values of $\lambda$. This is to be expected if we consider higher expansion rates to lead to smaller values of $Q$ and $J$. The sensitivity of this solution to the bias parameter $\rho$ is further illustrated in Appendix \ref{app2}.

Another interesting feature of this solution, which complements its full characterization, is the dependency of the sign of $\Delta$ on the value of $\rho$. This relation also depends on the value of $g$. In fact, we can rewrite the last restriction in \eqref{restrictions_A3} as:
\begin{equation}
    v_0\frac{\Delta}{Q_0} = \left(\rho - \rho_c\right) \left(g-1\right) \frac{Q_0^2 + J_0^2}{Q_0^2 J_0^2} \tilde{c}\frac{v_0}{\xi_0}\frac{2+Q_0^2+J_0^2}{2},
\end{equation}
where $\rho_c = \frac{J_0^2}{Q_0^2+J_0^2}$. Now it is clear how the sign of $\Delta$ varies with $\rho$ and $g$. This relation stems from the fact that both the charge and the current must evolve in the same way, in this case $\gamma=\delta=0$. The time variation interpretation can be useful here. In fact, if we consider $g>1$, then $Q$ and $J$ will be sensitive to loop production and their values will decrease with it. Further assuming that $Q_0^2=J_0^2$, we have $\rho_c=\frac{1}{2}$, which corresponds to the unbiased case. If $\rho >\frac{1}{2}$, then the charge will be less sensitive. This means the current will get smaller than the charge. For it not to happen, the magnetic field should influence $Q$ in a way that decreases its value. This happens precisely when $\Delta>0$. For different values of $Q_0^2$ and $J_0^2$, the value of $\Delta$ must adjust depending on the value of $\rho$ as well. The opposite happens in the $g<1$ regime. 

We can now consider the case where both the charge and the current grow with time. While this should not be an asymptotic scaling solution, since one expects a growing current to eventually saturate to a maximum possible value (at least if given a sufficient amount of time), the solution is nevertheless worth reporting, since it could in principle illustrate how a current-carrying string network might reach such a saturation point, and the role of the primordial magnetic field in this regard. Moreover, it is not guaranteed that the underlying assumption of a linear dependence on the chirality in \eqref{Witten_action} remains valid in this regime, where both the charge and the current continue to grow. Nevertheless, the solution yields
\begin{subequations}
	\begin{align}
		\alpha &= \lambda-p \\
		\gamma &= \delta = 1-\lambda + p \\
		\epsilon &= 1,
	\end{align}
\end{subequations}
subject to the restrictions
\begin{subequations}
		\begin{align}
			&\frac{g\tilde{c}}{2}\frac{v_0}{\xi_0} = -p \\
			&\frac{k_v}{\xi_0} = 2 \frac{Q_0 \Delta}{Q_0^2+J_0^2} \\
			&v_0\frac{k_v}{\xi_0}-\rho \left(g-1\right)\frac{\tilde{c}}{2}\frac{v_0}{\xi_0}\frac{Q_0^2+J_0^2}{J_0^2} = 1+p \\
			&v_0\frac{\Delta}{Q_0} = \left[\frac{\rho}{J_0^2}-\frac{1-\rho}{Q_0^2}\right]\left(g-1\right)\tilde{c}\frac{v_0}{\xi_0}\frac{Q_0^2+J_0^2}{2}.
		\end{align}
	\end{subequations}
Note that if $p=\lambda-1$ we get $\gamma=\delta=0$ and $\alpha=1$ as in the previous case. The first condition implies that $p<0$, which in turn implies $\lambda<1$, corresponding to neither the radiation nor the matter dominated eras. Nevertheless, this could in principle be a viable transient scaling solution in an early phase of cosmological kination, with $\lambda=1/2$. As $\rho$ remains undetermined, this set of restrictions has a very large parameter space to study. Therefore, we will analyze the familiar values of $\rho = 0$, $\rho = 1$ and the interesting case of $\rho = 1/2$.

For $\rho = 0$, the previous conditions would require that $g<0$, implying that the network would gain energy through the production of loops, which is unphysical. 

For $\rho=1$, we get $\frac{v_0}{\xi_0} = \frac{2}{k_v + \tilde{c}} \approx 1.3$. The ratio between the 'initial' averaged charge and current is $\frac{J_0^2}{Q_0^2} = \frac{\left(g-1 \right) \tilde{c}}{k_v}$, and must be positive, yielding $g>1$. These restrictions have implications on the possible rates of the expansion of the Universe. More specifically, $\lambda < \frac{k_v}{k_v + \tilde{c}} \approx 0.86$, again excluding the radiation and matter eras.

For $\rho =1/2$, several of the above results remain. We still have $\frac{v_0}{\xi_0} = \frac{2}{k_v + \tilde{c}} \approx 1.3$, and the charge to current ratio is $\frac{Q_0^2}{J_0^2} = \frac{2k_v}{\left(g-1\right) \tilde{c}}+1$, leading to $g>1-\frac{2k_v}{\tilde{c}}$. However, this value is negative and therefore we shall have $g>0$. This also means that $\lambda<1$, meaning higher expansion rates, relative to the case of $\rho=1$, are allowed. 

In these cases, where both charges and currents grow with time, we have $\eta>0$ and $\frac{E_0}{E} \rightarrow 0$, as it would be expected. This may explain why, in these cases, we were not able to determine $Q_0^2$ or $J_0^2$.

An overall look at the solutions of this first family reveals a relation between the dominant degree of freedom $\mathcal{D}_0^2$ and the bias parameter, $\rho$. In fact, in $A1$, we have $\mathcal{D}_0^2 = Q_0^2$ and $\rho = 0$; in $A2$, $\mathcal{D}_0^2 = J_0^2$ and $\rho = 1$ and in $A3$, $\mathcal{D}_0^2 = Q_0^2 + J_0^2$ and $0 \leq \rho \leq 1$. To explain this, we can imagine solution $A1$, where, although the charge does not grow, it dominates over the current, everywhere. The bias parameter measures whether the charge is more or less sensitive to loop production. If $\rho=0$, as it is the case for $A1$, then the current would feel no effect due to loops. But this is clearly expected in a situation where the current has no contribution to the overall total energy. In other words, the current does not feel any repercussions due to loop production, because there is no current. For $\eta > 0$, this analysis may not be straightforward, and the relation with the bias parameter may be more complex.

\subsection{Solutions with $\eta < 0$ and $\beta=0$}

There is only one solution satisfying $\eta < 0$ and $\beta=0$. We henceforth call it solution $B$ and it yields
\begin{subequations}
	\begin{align}
	\alpha &= \epsilon = 1 \\
	\beta &= 0\\
	\gamma &= 1-\lambda+p \\
	\delta &= 1-\lambda+p +\frac{v_0}{Q_0}\Delta.
\end{align}
\end{subequations}
As both charge and current decay over time, the energy of the network is given by the bare string energy, $\frac{E_0}{E} \rightarrow 1$. In other words, the network approaches the Nambu-Goto limit, with the magnetic field implying that charge and current decay at different rates.

It is important to highlight that this solution is only possible if $g=1$. This makes sense, because $g$ characterizes how the averaged charge and current are influenced by loop formation. For a vanishing charge and current, we couldn't have otherwise. In the same way, with no $Q$ and $J$, the bias parameter is ill-defined. And this also happens precisely when $g=1$.

We can also note that the evolution of the charge and current is related to the parameter $\Delta$. If $\Delta>0$, $\delta>\gamma$, and vice-versa. The restrictions also depend on the sign of $\Delta$. If it is positive, then we have
\begin{subequations}
		\begin{align}
		& \lambda > \frac{2}{1+\frac{\tilde{c}}{k_v}} \approx 1.7\\
		& v_0^2 = \frac{1}{1+\frac{\tilde{c}}{k_v}}\frac{1}{\lambda}<\frac{1}{2} \\
		& \frac{v_0}{\xi_0} = \frac{2}{k_v + \tilde{c}} \approx 1.3\\
		&p = 1-\frac{2\tilde{c}}{k_v + \tilde{c}}-\frac{v_0}{Q_0}\Delta.
	\end{align}
	\end{subequations}
The evolution of the magnetic field cannot be fully determined. On the other hand, we keep the previous value of $\frac{v_0}{\xi_0}$, and ultra-relativistic velocities are not allowed. However, the most notable feature is that only large enough expansion rates are allowed, as if they are needed in order for the charge and current to dilute away. This limiting expansion rate is determined by the loop chopping efficiency: in the limit $\tilde{c}\to0$ the limit is $\lambda>2$, i.e., as long as loop production occurs this solution is possible in the matter era. For $\Delta<0$, the restrictions are the same with the exception of $\lambda> \frac{2}{1+\frac{\tilde{c}}{k_v}} - \frac{v_0}{Q_0} \Delta > \frac{2}{1+\frac{\tilde{c}}{k_v}} \approx 1.7$, yielding, in the end, the same behavior. 

\subsection{Solutions with $\beta < 0$}

This last family is the most populated one, with a total of 6 different solution branches. As has already noticed, for a decaying velocity we need a constant dominant degree of freedom, $\eta=0$. However, for the growing power term to vanish in the velocity equation, we need $\frac{L_0^2}{\xi_0^2}\mathcal{D}_0^2 = 1$. On the other hand, $\eta=0$ implies that $\frac{L_0^2}{\xi_0^2} = \frac{2}{2+\mathcal{D}_0^2}$, which can only be reconciled if $\mathcal{D}_0^2 = 2$. This value is very specific and corresponds to an equal partition of the total energy between string contribution and the one from charges and currents. In other words, in this family, $\frac{E_0}{E} = 1/2$. This is very interesting, because in family $A$, the energy fraction values for a vanishing velocity, $v_0^2 \rightarrow 0$, would match this ratio.

The first solution branch, denoted $C1$, reads
\begin{subequations}
	\begin{align}
	\alpha &= \epsilon = \frac{\lambda}{2} + \frac{g\tilde{c}}{2}\frac{v_0}{\xi_0}\\
	\beta &= \gamma = \alpha -1\\
	\delta &= 0,
\end{align}
\end{subequations}
with the restrictions
\begin{subequations}
	\begin{align}
	&\rho=1\\
	&\frac{g\tilde{c}}{2}\frac{v_0}{\xi_0} = 1-\frac{3}{2}\lambda + \frac{Q_0 \Delta}{2 v_0}\\
	&\frac{v_0}{\xi_0} = \frac{\lambda}{k_v-\left(g-1\right)\tilde{c}}\\
	&v_0\frac{\Delta}{Q_0} = \left(k_c-\frac{g\tilde{c}}{2}\right)\frac{v_0}{\xi_0}+\frac{3}{2}\lambda\\
	&\left[1+\frac{g\tilde{c}}{k_v-\left(g-1\right)\tilde{c}}\right]\lambda < 2.
\end{align}
\end{subequations}

\begin{figure}
	\centering
	\includegraphics[width=\columnwidth]{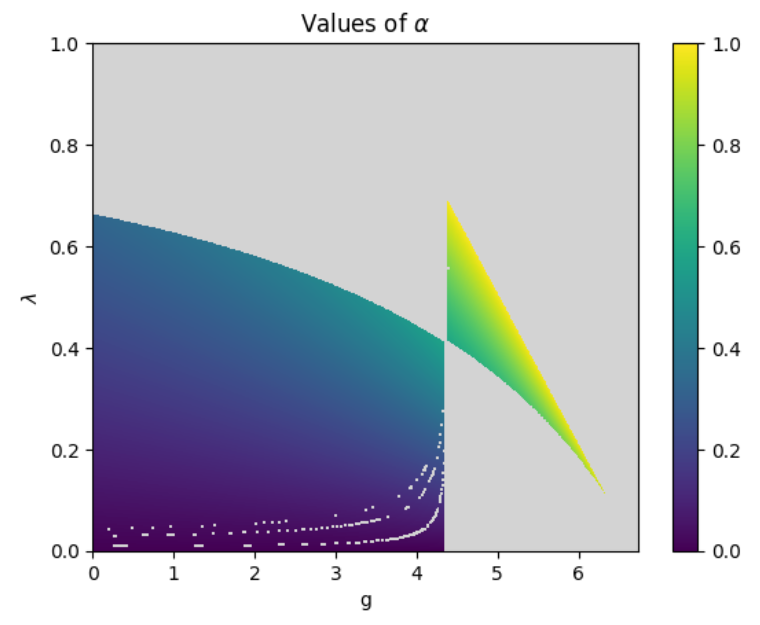}
	\caption{Possible values of $\alpha=\epsilon$ considering the solution $C1$. Regions in gray do not belong to the parameter space.}
	\label{fig:C1_alpha}
\end{figure}

The second restriction implies that $g<1+\frac{k_v}{\tilde{c}}$, while the last one leads to $g<\left(1+\frac{k_v}{\tilde{c}}\right)\left(2-\lambda\right)$. Accounting for a positive value of $g$, we get an upper limit for the expansion rate of the Universe, $\lambda<2$. However, in \autoref{fig:C1_alpha} we see that the parameter space only allows smaller values of $\lambda$.  We can also see the possible values of $\alpha$, which characterizes the evolution of the characteristic length, and, as it is expected, it varies between $0$ and $1$.

\begin{figure}
	\centering
	\includegraphics[width=\columnwidth]{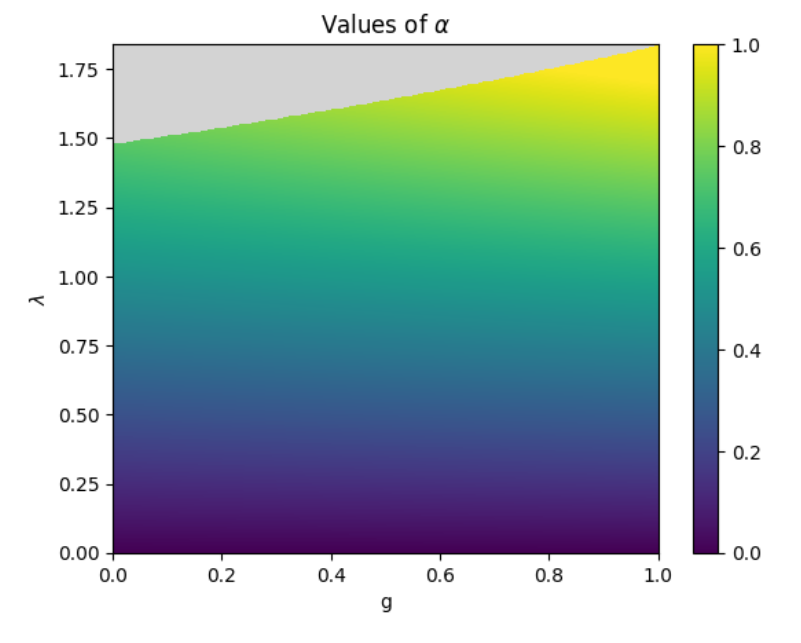}
	\caption{Possible values of $\alpha=\epsilon$ considering the solution $C2$. Regions in gray do not belong to the parameter space}
	\label{fig:C2_alpha}
\end{figure}

Moving to the second solution, denoted $C2$, this has the form
\begin{subequations}
	\begin{align}
	\alpha &= \epsilon = \frac{\lambda}{2} + \frac{g\tilde{c}}{2}\frac{v_0}{\xi_0}\\
	\beta &= \alpha -1\\
	\gamma &= \left(g-1\right)\tilde{c}\frac{v_0}{\xi_0}\\
	\delta &=0.
\end{align}
\end{subequations}
We can already see that in order for the charge to decay, we must have $g<1$. The full set of restrictions are
\begin{subequations}
	\begin{align}
	&\rho=1\\
	&\frac{v_0}{\xi_0} = \frac{\lambda}{k_v}\\
	&\frac{Q_0 \Delta}{v_0} = 3\lambda-2+g\tilde{c}\frac{v_0}{\xi_0}\\
	&\left[1+\frac{g\tilde{c}}{2 k_v}-2\left(g-1\right)\frac{\tilde{c}}{k_v}\right]\lambda<2.
\end{align}
\end{subequations}
By solving the last inequality for $g$, we get the following upper bound on the expansion of the Universe, $\lambda<\frac{4}{2+\frac{\tilde{c}}{k_v}}\approx 1.84$. Therefore, this solution is also not possible in the matter dominated era. As before, we varied the parameter $g$ and $\lambda$ to find the parameter space, and to determine the values of $\alpha$. The results are presented in \autoref{fig:C2_alpha}.

The next solution, $C3$, has the simpler form
\begin{subequations}
	\begin{align}
	\alpha &= \epsilon = 1-\lambda \\
	\beta &= -\lambda\\
	\gamma &= -v_0\frac{\Delta}{Q_0}+\left(g-1\right)\tilde{c}\frac{v_0}{\xi_0}\\
	\delta &= 0,
\end{align}
\end{subequations}
with the constraints
\begin{subequations}
	\begin{align}
	&\rho = 1\\
	&\frac{v_0}{\xi_0} = \frac{2}{3 k_v +3\tilde{c}-2g\tilde{c}}\\
	&\lambda = 2\frac{k_v+\tilde{c}-g\tilde{c}}{3 k_v +3\tilde{c}-2g\tilde{c}}\\
	&p=2\lambda-1-v_0\frac{\Delta}{Q_0}+\left(g-1\right)\tilde{c}\frac{v_0}{\xi_0}\\
	&\frac{\Delta}{Q_0}>\frac{k_v}{\xi_0}.
\end{align}
\end{subequations}
 Here we can see that $\lambda$ depends on the value of $g$, therefore reducing the dimension of the parameter space. In order to have $\lambda>0$, we must have $g<1+\frac{k_v}{\tilde{c}}$, which ensures that $\frac{v_0}{\xi_0}>0$, as it should be. Moreover, as it already seems a feature of this family of solutions, an upper bound can be found for the expansion rate. In fact, we must have $\lambda<\frac{2}{3}$, leading to $\frac{1}{3}<\alpha <1$. 

Until now, all the C-class solutions featured $\rho=1$, and, as it should be expected by now, $\delta>\gamma$. There are three solutions left to be discussed, with different behaviors: $C4$ has $\rho=0$ and $\gamma>\delta$, while $C5$ and $C6$ feature $\gamma=\delta$ and $\rho = \frac{J_0^2}{J_0^2 + Q_0^2}$.

\begin{figure}
	\centering
	\includegraphics[width=\columnwidth]{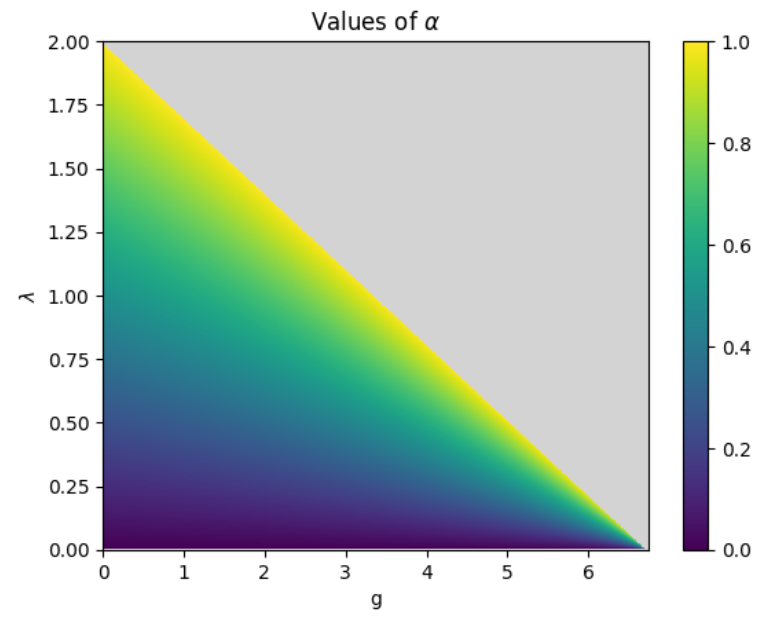}
	\caption{Possible values of $\alpha=\epsilon$ considering the solution $C4$. Regions in gray do not belong to the parameter space}
	\label{fig:C4_alpha}
\end{figure}

Solution $C4$ reads
\begin{subequations}
	\begin{align}
	\alpha &= \epsilon = \frac{\lambda}{2} + \frac{g\tilde{c}}{2}\frac{v_0}{\xi_0}\\
	\beta &= \alpha -1\\
	\gamma &= 0\\
	\delta &= \left(g-1\right)\tilde{c}\frac{v_0}{\xi_0}.
\end{align}
\end{subequations}
This solution seems a mirror image of $C2$, with the change between charge and current, and of some of the restrictions, which become
\begin{subequations}
	\begin{align}
	&\rho=0\\
	&\frac{v_0}{\xi_0} = \frac{\lambda}{k_v-\left(g-1\right)\tilde{c}}\\
	&\frac{Q_0 \Delta}{v_0} = 3\lambda-2+g\tilde{c}\frac{v_0}{\xi_0}\\
	&p = \frac{3}{2}\lambda-2+ \frac{g\tilde{c}}{2}\frac{v_0}{\xi_0}\\
	&\left[1+\frac{g\tilde{c}}{k_v-\left(g-1\right)\tilde{c}}\right]\lambda<2.
\end{align}
\end{subequations}
The last condition implies $g<\left(1+\frac{k_v}{\xi_0}\right)\left(1-\frac{\lambda}{2}\right)$. Accounting for a positive value of $g$, we get $\lambda<2$. Again,  the range of possible values of $\alpha$ is represented in \autoref{fig:C4_alpha}.

Finally, we address the solutions with an equal scaling behavior for the charge and current. But before, we should discuss the particular, specific value of the bias parameter, $\rho = \frac{J_0^2}{J_0^2 + Q_0^2}$. This value coincides precisely with $\rho_c$ studied in solution $A3$, for $\eta=0$. Although $\gamma = \delta$, we can still have the string network dominated by the charge, if $Q_0^2 \gg J_0^2$, and vice-versa. If that happens, then we would have $\rho \rightarrow 0$, as it should be expected. In the same way, for $J_0^2 \gg Q_0^2$, we have $\rho \rightarrow 1$. This is a confirmation of our previous analysis.

\begin{figure}
	\centering
	\includegraphics[width=\columnwidth]{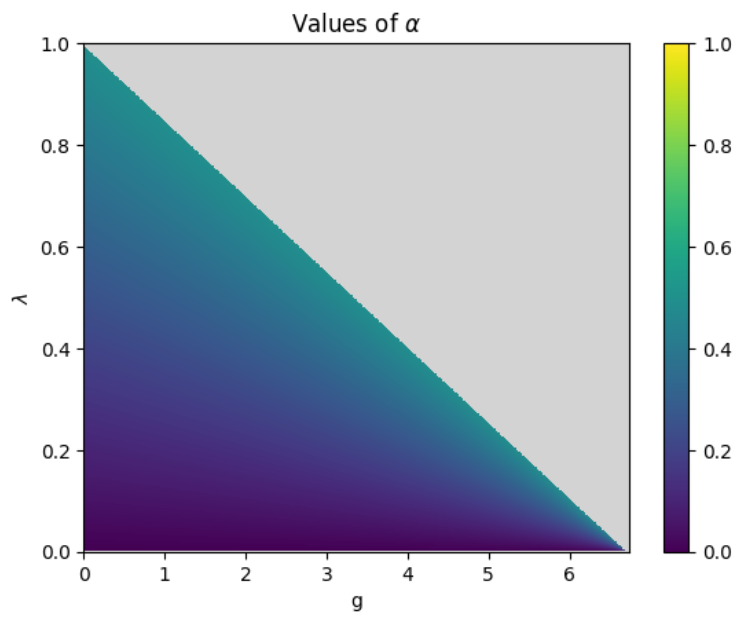}
	\caption{Possible values of $\alpha=\epsilon$ considering the solution $C5$. Regions in gray do not belong to the parameter space}
	\label{fig:C5_alpha}
\end{figure}

In solution $C5$ we have
\begin{subequations}
	\begin{align}
	\alpha &= \epsilon = \frac{\lambda}{2}+\frac{g\tilde{c}}{2}\frac{v_0}{\xi_0}\\
	\beta &= \alpha -1\\
	\gamma &= \delta = 0,
\end{align}
\end{subequations}
which again seems an extension of solutions $C2$ and $C4$. The restrictions are
\begin{subequations}
	\begin{align}
	&\rho = \frac{J_0^2}{J_0^2+Q_0^2}\\
	&\frac{v_0}{\xi_0} = \frac{\lambda}{k_v-\left(g-1\right)\tilde{c}}\\
	&\frac{Q_0 \Delta}{v_0} = 3\lambda-2+g\tilde{c}\frac{v_0}{\xi_0}\\
	&p = \frac{3}{2}\lambda-2+ \frac{g\tilde{c}}{2}\frac{v_0}{\xi_0}\\
	&\left[1+\frac{g\tilde{c}}{k_v-\left(g-1\right)\tilde{c}}\right]\lambda<1.
\end{align}
\end{subequations}
The values of $J_0^2$ and $Q_0^2$ remain undetermined, as does the numerical value of the bias parameter. As $g$ must be positive, we have $\lambda<1$. The value of $\alpha$ is represented in \autoref{fig:C5_alpha}, and we can see that it cannot reach such high values as before. This is a consequence of the last restriction, which can be rewritten as $\alpha < 1/2$.

The last solution, $C6$, reads 
\begin{subequations}
	\begin{align}
	\alpha &= \epsilon = \frac{\lambda}{2}\\
	\beta &= -1-\frac{\lambda}{2}\\
	\gamma &= \delta = 0,
\end{align}
\end{subequations}
with the constraints
\begin{subequations}
	\begin{align}
	&\rho = \frac{J_0^2}{J_0^2+Q_0^2}\\
	&\frac{v_0}{\xi_0} = \frac{\lambda}{\left(g-1\right)\tilde{c}}\\
	&\frac{Q_0 \Delta}{v_0} = \lambda - 2\\
	&p = \frac{\lambda}{2}-2\\
	&\lambda<2.
\end{align}
\end{subequations}

In order to have $\frac{Q_0}{v_0}>0$ and $\frac{v_0}{\xi_0}>0$, we must have $\Delta<0$ and $g>1$, respectively. In this situation the parameter space corresponds to the area given by $0<\lambda<2$ and $g>1$, and the value of $\alpha$ is trivially determined.

It should be noted that, in this family, the previous analysis of the behavior of $g$ as a function of $\lambda$ no longer holds. This is due to the nonexistence of a constant critical expansion rate. In fact $\lambda_c \propto \frac{v_0}{\xi_0}$, and now we have $\frac{v_0}{\xi_0}=\frac{v_0}{\xi_0}\left( g, \, \lambda\right)$. So for a given value of $\lambda$, we can have $g>1$ and $g<1$, as it can be seen in several solutions in this family.

\section{\label{section4}Conclusions}

We have phenomenologically explored a physical mechanism under which a current-carrying cosmic string network could conceivably gain energy, instead of only losing it. Building upon the previously developed CVOS formalism, we have extended it in order to include the influence of an external magnetic field, assumed to exist on a sufficiently large (mesoscopic) scale such that it can be described as purely time-dependent one---in other words, such that spatial dependencies can be averaged out. The corresponding set of macroscopic equations is given by \ref{Equations_CVOS_Magnetic}. A first look indicates that, depending on the sign of the parameter $\theta$ (or, equivalently, $\Delta$), the magnetic field can have a positive or a negative effect on charge growth, respectively. This shows the possible dual effect of a magnetic field, as one might have anticipated.

Assuming power-law dependencies for the CVOS-like model degrees of freedom (characteristic and correlation length, and averaged velocity, charge and current) as well as for the cosmological scale factor and magnetic field time dependence, we have searched for physically acceptable asymptotic scaling solutions, and have classified them into three different classes. An organized summary of the main aspects of all ten solutions solutions can be found in \autoref{only_table}. Although these solutions were numerically studied for specific values of the VOS model parameters $k_v$ and $\tilde{c}$ (their value choices being backed by numerical simulations), only one of the ten individual solutions, namely $A3$, allowed for the network to asymptotically continue to gain energy, in the form of charge and current.

\begin{table*}
    \centering
	\begin{tabular}{|c|c|c|c||c||c|c|c|c|c|c|} 
		\hline
		Exponent & $A1$ & $A2$ & $A3$ & $B$ & $C1$ & $C2$ &$C3$&$C4$&$C5$&$C6$\\
		\hline
		$\alpha$ & 1 & 1& $\left[0, 1\right]$&1& $\left[0, 1\right]$&$\left[0, 1\right]$&$\left[\frac{1}{3}, 1\right]$&$\left[0, 1\right]$&$\left[0, \frac{1}{2}\right]$&$\left[0, 1\right]$\\
		$\epsilon$& $=\alpha$ & $=\alpha$ &1&$=\alpha$&$=\alpha$&$=\alpha$&$=\alpha$&$=\alpha$&$=\alpha$&$=\alpha$ \\ 
		\hline
		$\beta$& 0 & 0& 0& 0&$<0$&$<0$&$<0$&$<0$&$<0$&$<0$ \\ 
		\hline
		$\gamma$&0&$\left]-2.9, 0\right[$&$\geq0$&$<0$&$<0$&$<0$&$<0$&0&0&0\\
		$\delta$&$\approx -0.3$&0&$=\gamma$&$<0$&0&0&0&$<0$&$=\gamma$&$=\gamma$\\
		\hline
		$\lambda$&-&-&$<1$&$\gtrsim1.7$&$\lesssim 0.8$& $\lesssim 1.84$&$<\frac{2}{3}$&$<2$&$<1$&$<2$\\
		\hline
	\end{tabular}
	\caption{Summary of the most important scaling features of all the solutions. The possible values of $\lambda$ in the solutions $A1$ and $A2$ were not studied. The $\lambda$ limit in the $A3$ column corresponds to the case where $\eta>0$.}
	\label{only_table}
\end{table*}

One important restriction on our analysis of the scaling solutions and their behavior resulted from the linear dependence on $k$, on the action, cf. Eq. (\ref{Witten_action}). This is expected to be a good approximation, since in most circumstances the charges and currents are expected to be small. Nevertheless, it is possible, though only in quite narrow regions of the model parameter space, that the charge and current can grow, in which case the linear ansatz might not be justified. In that case, it is also conceivable that backreaction effects can be significant.  Last but not least, a more general function, $f\left( k \right)$, could possibly lead to additional solution branches. An exploration of this scenario can be a future line of work.

Finally, we note that the VOS-type models, although quantitative and self-contained, are only as reliable as the simulations with which their model parameters are calibrated. As previously mentioned, for the CVOS model, this process is currently ongoing, with initial results reported in \cite{Correia24}. It would therefore be desirable to also explore the scaling solutions presented herein through numerical simulations. Conceptually, it is possible to extent such a current-carrying string field theory code to include a simple background magnetic field, e.g. one aligned with one of the axes of the simulation box, although a practical implementation remains to be attempted, and is left for future work.

\appendix
\section{\label{app1}Determination of the $\theta$ evolution equation}

To determine the evolution equation for $\theta$, we must compute the time derivative of its microscopic counterpart and then average it. However, firstly, we shall construct the trihedron which will allow us to have a better description of the microscopic structure of the strings. We define
\begin{subequations}
    \begin{align}
        \hat{e}_0 &= \frac{\dot{\mathbf{x}}}{\sqrt{\dot{\mathrm{x}}^2}} \\
        \hat{e}_1 &= \frac{\mathbf{x}'}{\sqrt{\mathrm{x}'^2}} \\
        \hat{e}_2 &= \frac{\dot{\mathbf{x}} \times\mathbf{x}'}{\sqrt{\dot{\mathrm{x}}^2\mathrm{x}'^2}}.
    \end{align}
\end{subequations}
We should note that this trihedron is only defined on points where the velocity $\dot{\mathbf{x}}$ is not zero and where $\mathbf{x}'$ is not ill-defined (as it happens in a kink). We can now write $\theta = -\frac{1}{B}\langle \mathbf{B}\cdot \hat{e}_2\rangle$. When computing its time derivative, we shall not worry about time variations along the direction of the magnetic field. Although they are relatively easy to implement (at the expense of yet another macroscopic quantity), we shall not consider it here.

From the variation of the action in Eq. \ref{Witten_action}, one of the resulting equations describes the time evolution of the velocity,
\begin{align}
    \epsilon \tilde{U} \ddot{\mathbf{x}} &+ \frac{\dot{a}}{a}\epsilon \left( \tilde{U} + \tilde{T}\right)\left(1-\dot{\mathrm{x}}^2\right) \dot{\mathbf{x}}-\frac{1}{a^2}\frac{\partial}{\partial \tau}\left( a^2 \Phi\right)\\ &-\frac{\partial}{\partial \sigma}\left( \frac{\tilde{T}}{\epsilon} \right) \mathbf{x}'-2\Phi \dot{\mathbf{x}}' - \frac{\tilde{T}}{\epsilon}\mathbf{x}'' = -e\mathbf{B}\times \left(\phi' \dot{\mathbf{x}} - \dot{\phi} \mathbf{x}'\right)\notag
\end{align}
If we are not considering changes on the magnetic field direction, we must only worry with the variations of $\hat{e}_2$. As it is an unit vector any variation should be along $\hat{e}_0$ and $\hat{e}_1$. So
\begin{align}
    \frac{\partial}{\partial \tau} \left(\hat{n}\cdot \hat{e}_2 \right) &= \hat{n} \cdot\left[ \left( \frac{\partial \hat{e}_2}{\partial \tau}\cdot\hat{e}_0\right)\hat{e}_0 + \left( \frac{\partial \hat{e}_2}{\partial \tau}\cdot\hat{e}_1\right)\hat{e}_1\right]
\end{align}
The two components of $\frac{\partial \hat{e}_2}{\partial \tau}$ can be computed through $\frac{\partial \hat{e}_2}{\partial \tau}\cdot\hat{e}_0=-\frac{\partial \hat{e}_0}{\partial \tau}\cdot\hat{e}_2$ and $\frac{\partial \hat{e}_2}{\partial \tau}\cdot\hat{e}_1=-\frac{\partial \hat{e}_1}{\partial \tau}\cdot\hat{e}_2$. We then have 
\begin{align}
    \frac{\partial}{\partial \tau}& \left(\hat{n}\cdot \hat{e}_2 \right) = - 2\frac{\Phi}{\epsilon \tilde{U}}\frac{\dot{\mathbf{x}}'\cdot \hat{e}_2}{\sqrt{\dot{\mathrm{x}}^2}}\left(\hat{n}\cdot\hat{e}_0\right) - \frac{\tilde{T}}{\epsilon^2 \tilde{U}}\frac{\mathbf{x}''\cdot \hat{e}_2}{\sqrt{\mathrm{x}'^2}}\left(\hat{n}\cdot\hat{e}_0\right) \\
    &- \frac{ea}{\epsilon \tilde{U}}\sqrt{\mathrm{x}'^2}j\left(\mathbf{B}\cdot\hat{e}_1\right)\left(\hat{n}\cdot\hat{e}_0\right) \notag \\
    &- \frac{ea}{\epsilon \tilde{U}}\sqrt{1-\dot{\mathrm{x}}^2}\sqrt{\frac{\mathrm{x}'^2}{\dot{\mathrm{x}}^2}}q\left(\mathbf{B}\cdot\hat{e}_0\right)\left(\hat{n}\cdot\hat{e}_0\right) - \frac{\dot{\mathbf{x}}'\cdot \hat{e}_2}{\sqrt{\mathrm{x}'^2}}\left(\hat{n}\cdot\hat{e}_1\right) \notag
\end{align}
By averaging the previous equation according to Eq. \ref{averageq} and assuming the linear behavior of the function $f$, we get the evolution equation of Eq. \ref{theta_evolution}. We should note that, of the five new macroscopic parameters introduced, only three are related to the magnetic field. So, in principle, similarly to the momentum parameter, $k_v$, both $k_1$ and $k_2$ can be studied and analyzed through numerical simulations without an external magnetic field. However, it should be noted that this kind of analysis presents some difficulties, and, while not impossible, it would be easier to determine the parameter $\theta$ via numerical simulations (even though it would demand the extra caveat of incorporating an external magnetic field into the simulations). 

Similar procedures can be applied to $\theta_0$, $\theta_1$ and $\nu$. In fact, we have
\begin{subequations}
    \begin{align}
        \frac{d \theta_0}{d\tau} &= \frac{B\theta_1}{W^2}\frac{\sqrt{1-v^2}}{v}\left[ 2\frac{\dot{a}}{a}QJ + \dot{Q}J + Q\dot{J} \right] \\
        &- 2\frac{QJ}{W^2}\sqrt{1-v^2}\frac{k_v}{\xi}B \theta_1 + 2\frac{QJ}{W^2}\frac{1-v^2}{v}k_1 B \theta \notag \\&+ \left[ 1+\frac{Q^2+J^2}{W^2}\right] \frac{1-v^2}{v}k_2 B \theta - \frac{eaQ}{W^2}\frac{1-v^2}{v}B^2\theta \theta_0 \notag 
    \end{align}
\end{subequations}
\begin{subequations}
    \begin{align}
        \frac{d\theta_1}{d\tau} &= -\frac{B\theta_1}{W^2}\frac{\sqrt{1-v^2}}{v}\left[ 2\frac{\dot{a}}{a}QJ + \dot{Q}J + Q\dot{J} \right]- k_1B\theta\\&+2\frac{QJ}{W^2}\sqrt{1-v^2}\frac{k_v}{\xi}B \theta_1-\frac{eaJ}{W^2}\sqrt{1-v^2}B^2\theta \theta_1 \notag
    \end{align}
\end{subequations}
\begin{subequations}
    \begin{align}
        \frac{d\nu}{d\tau} &= 2\frac{B^2\theta_0\theta_1}{W^2}\frac{\sqrt{1-v^2}}{v}\left[ 2\frac{\dot{a}}{a}QJ + \dot{Q}J + Q\dot{J} \right] \\
        &- 4\frac{QJ}{W^2}\sqrt{1-v^2}\frac{k_v}{\xi}B^2\theta_0 \theta_1 + 4\frac{QJ}{W^2}\frac{1-v^2}{v}k_1 B^2 \theta \theta_0\notag \\&+ 2\left[ 1+\frac{Q^2+J^2}{W^2}\right] \frac{1-v^2}{v}k_2 B^2 \theta \theta_0 - 2\frac{eaQ}{W^2}\frac{1-v^2}{v}B^3\theta \nu. \notag
    \end{align}
\end{subequations}
These form a set of equations which allow us to compute the time evolution of the parameters related to the magnetic field. It should be noted that they form a closed system, as their time derivatives do not depend on any additional macroscopic parameter.

\section{\label{app2}Sensitivity of solution A3 of the bias parameter $\rho$.}

\begin{figure*}[htbp]
    \centering
    \includegraphics[width=\textwidth]{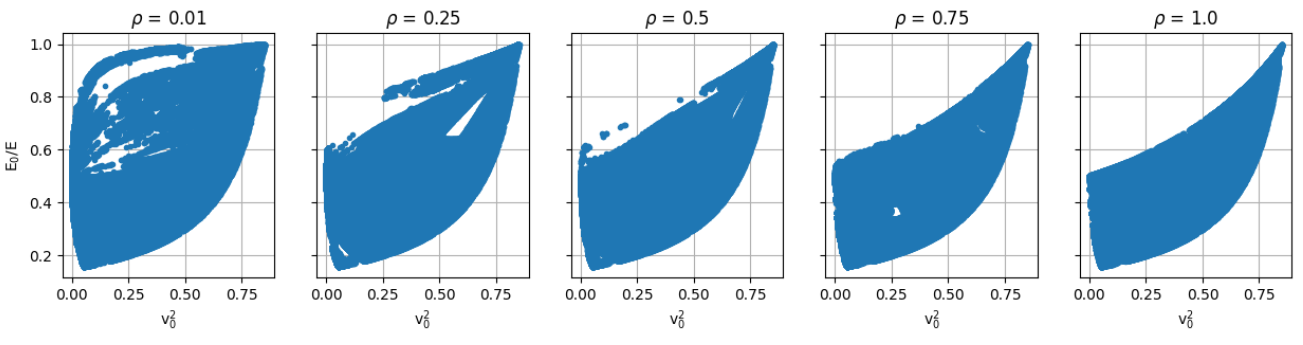}
    \caption{Bare string energy ratio as a function of the bias parameter, $\rho$, in the radiation dominated era, $\lambda=1$, considering solution $A3$ with $\eta=0$.}
    \label{different_rho_Rad}
\end{figure*}

\begin{figure*}[htbp]
    \centering
    \includegraphics[width=\textwidth]{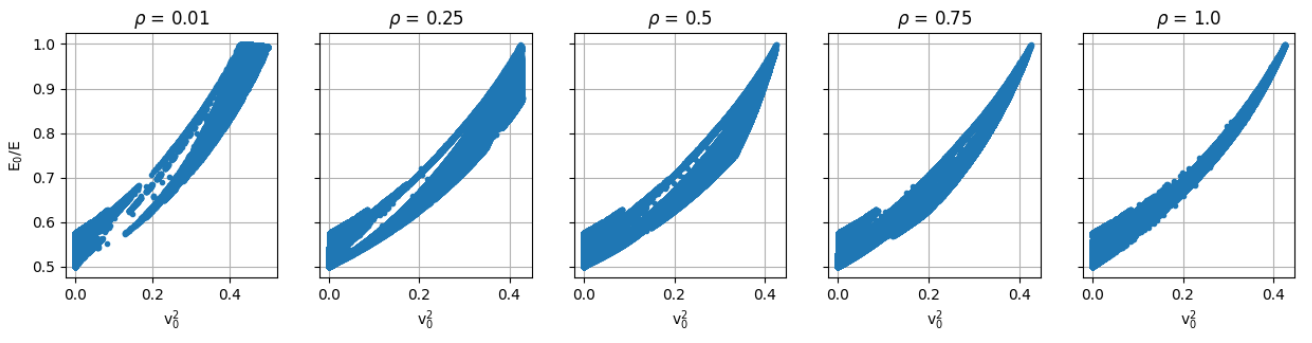}
    \caption{Bare string energy ratio as a function of the bias parameter, $\rho$, in the matter dominated era, $\lambda=2$, considering solution $A3$ with $\eta=0$.}
    \label{different_rho_Mat}
\end{figure*}

In \autoref{different_rho_Rad} and \autoref{different_rho_Mat} are represented the energy fraction values as a function of squared velocity for different values of the bias parameter, considering the radiation and matter dominated eras, respectively. We note that, for $\lambda=1$, the possible values have a broader range for a given velocity. In both eras, as $\rho$ increases, the energy fraction values have a narrower and a more consistent profile. It should also be highlighted that, in the limit of $v_0^2 \rightarrow 0$, we have $\frac{E_0}{E}\rightarrow \frac{1}{2}$, although, in order to appropriately see it, a zoom in is necessary.

%%%%%%%%%%%%%%%%%%%%%%%%%%%%%%%%%%%%%%%%%%%%%%%%%%%%%%%%%%%%%%%%%%%%%%%%%

\begin{acknowledgments}
This work was financed by Portuguese funds through FCT (Funda\c c\~ao para a Ci\^encia e a Tecnologia) in the framework of the project 2022.04048.PTDC (Phi in the Sky, DOI 10.54499/2022.04048.PTDC). CJM also acknowledges FCT and POCH/FSE (EC) support through Investigador FCT Contract 2021.01214.CEECIND/CP1658/CT0001 (DOI 10.54499/2021.01214.CEECIND/CP1658/CT0001). 
\end{acknowledgments}
 
\bibliography{article}

\end{document}